\documentclass[12pt]{article}

\usepackage{graphicx}
\usepackage{color}
\usepackage{mathptmx}      
\usepackage{amssymb}
\usepackage{amsmath}
\usepackage{ifsym}
\usepackage{sidecap}
\usepackage[colorlinks, linkcolor=blue, citecolor=blue, urlcolor=blue]{hyperref}

\usepackage{amssymb}

\begin{document}

\noindent
{\Large Estimate of sizes of small asteroids (cosmic bodies) by the method of stroboscopic radiolocation\footnote{\sl Accepted for publication in Acta Astronautica}}

\bigskip
\bigskip

\noindent
{\large V. D. Zakharchenko\footnote{\sl E-mail address: zakharchenko\_vd@mail.ru},  I. G. Kovalenko\footnote{\sl E-mail address: ilya.g.kovalenko@gmail.com}
O. V. Pak \footnote{\sl  E-mail address: olpak1@mail.ru}
)

\bigskip
\noindent
{\it Volgograd State University,
         Universitetskij Pr., 100,  Volgograd 400062, Russia
}

\begin{abstract}

Radiolocation methods of probing minor celestial bodies (asteroids) by the nanosecond pulses can be used for monitoring of near-Earth space with the purpose of identification of hazardous cosmic objects able to impact the Earth.

Development of the methods that allow to improve accuracy of determining the asteroid’s size  (i.e. whether it measures tens or hundreds meters in diameter) is important for correctly estimating the degree of damage which they can cause (either regional or global catastrophes, respectively). In this paper we suggest a novel method of estimating the sizes of the passive cosmic objects using the radiolocation probing by ultra-high-resolution nanosecond signals to obtain radar signatures. The modulation envelope of the reflected signal, which is a radar portrait of the cosmic object, is subjected to time scale transformation to carrier Doppler frequency by means of radioimpulse strobing. The shift of a strobe within the probing period will be performed by radial motion of the object which will allow to forgo the special autoshift circuit used in the oscillographic technical equipment.

The measured values of duration of radiolocation portrait can be used to estimate the mean radius of the object by using the average spatial length of the portrait. The method makes it possible to appraise the sizes of cosmic objects through their radiolocation portrait duration, with accuracy that is independent of the object’s range.

 \end{abstract}


\section{Introduction}\label{Introduction}

Radiolocation methods of probing of passive cosmic objects (large meteors and asteroids) can be used for surveying the near-Earth space for the purpose of recognition of objects that present danger upon impact with the Earth.

It is known that cosmic objects smaller than 10 meters in size do not reach Earth’s surface, burning up in  the atmosphere \cite{torinoscale}, and thus are not dangerous for the planet’s population. The bodies that are tens meters across are able to explode and cause serious destruction, while the objects with a size of hundreds meters in extent and larger would lead to a regional or global disaster. With that, the bodies ranging specifically from 70 to 200 meters in diameter present the maximum danger for the humanity in its characteristic timescale, since they have greater probability of impacting the Earth than the larger bodies and their average destructive effect is maximal (NASA NEO STD Report  \cite{NASA, morrison05}). Thus the questions of improving accuracy for estimating sizes of cosmic bodies crossing the Earth’s orbit are relevant even at present and the interest in them will only increase.

The shortcomings of the optical methods for measurement of linear dimensions of celestial bodies are that error increases proportionally with distance to the measured object. The reason for that is that the optical systems of measurement are actually angular observations, and consequently the errors in angle measurement result in errors of estimating diameters proportional to the monitored object’s distance. Besides, all optical means of ground-based observation are subject to errors due to atmospheric opacity and turbulence. Radar probing methods are free from the aforementioned drawbacks, their resolution is determined by the properties of the signals used and does not depend on distance.

The location of the radar systems is suggested at the geostationary orbit for the permanent monitoring of potentially dangerous directions. The advantage of this location is the absence of atmosphere noise which enhances the likelihood of  detection of  hazardous objects.

\section{Radiolocation portrait of an asteroid}\label{Radioportrait}

   \begin{figure}
   \includegraphics[width=1\textwidth]{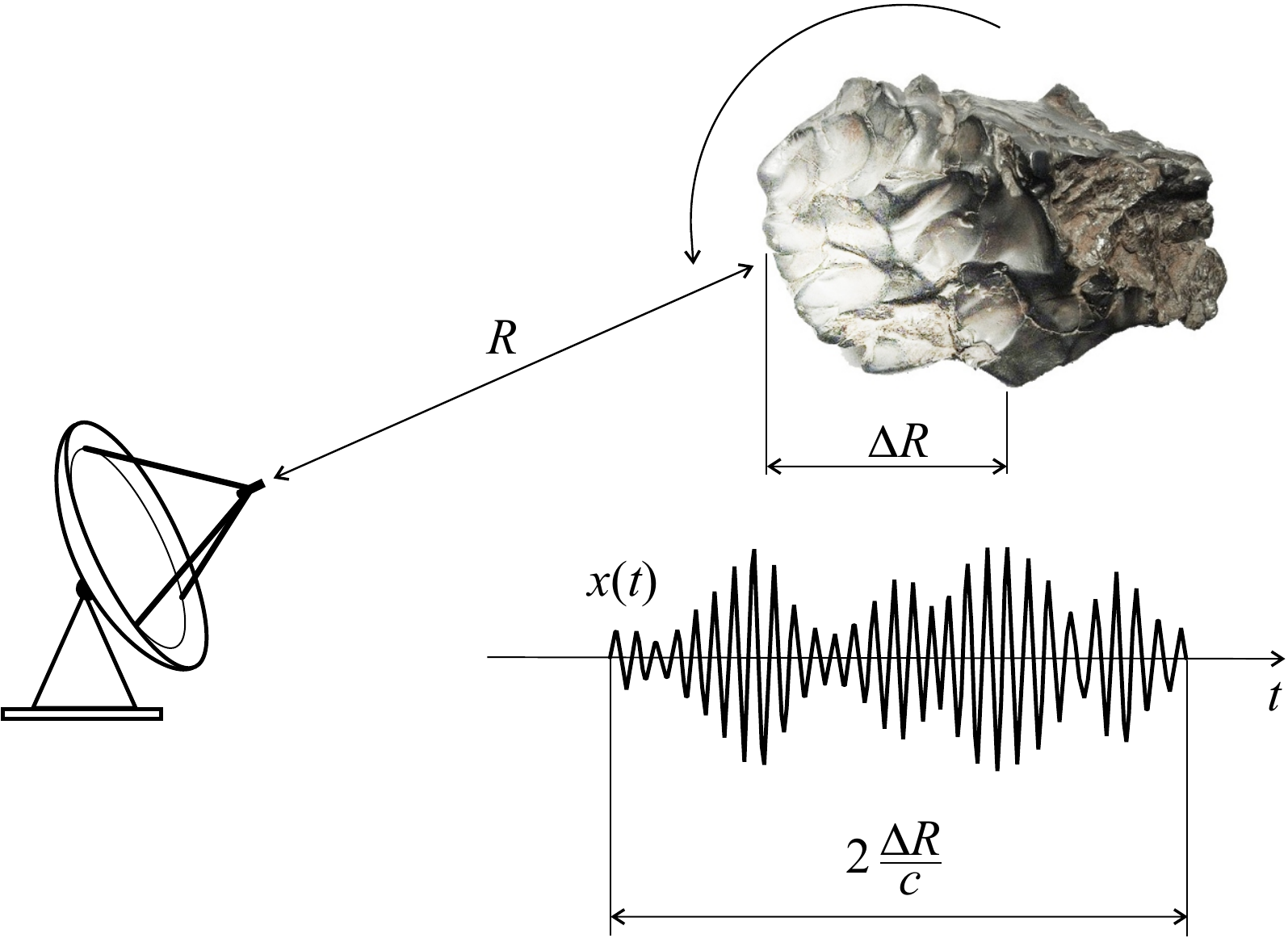} 
   \caption{The radar portrait-forming network.}
              \label{Fig1}%
    \end{figure}

The longitudinal asteroid’s dimension ($10 \div 100$ m) is determined by the radar signature duration with the use of  ultrashort ($\sim 3$ ns) RF pulses providing range resolution $\sim 0.5$ m.

The radar systems using the traditional narrow-band long-duration signals do not enable us to estimate the linear dimensions of the cosmic objects with the desired precision due to insufficient resolution. For the foregoing purposes one needs probing by high-resolution signals. In radiolocation the signals with large absolute width of spectrum  $\Delta f$, are defined as “highresolution” ones if they have high distance resolving power $\Delta r \approx 2c/\Delta f \ll L$ where $c$ is the speed of light and $L$ is the characteristic dimensions of the object reflecting signal \cite{lazorenko}. At that, the value $c\tau_u$, where $\tau_u$ is the signal duration, has the meaning of the spatial length of the signal. These signals produce the radiolocation portrait of the object, that is, the response $x(t)$ to high-resolution signal $x_0(t)$ that is governed by the radial dimension $\Delta R$ of the illuminated side of the object (Fig.~\ref{Fig1}).

The radiolocation portrait represents a target echo signal upon the condition of ``superresolution'' when the radar's distance resolution $\Delta r$ is much less than the linear dimensions of the target.
The process of transformation of the probing signal $x_0(t)$ to the reflected one $x(t)$ can be described by the integral
   \begin{equation}
      x(t) = \int_{-\infty}^{\infty} x_0(t^{\prime})h(t-t^{\prime}) d t^{\prime}
   \end{equation}
with the kernel as the aggregate of ``bright points'', that is, the local surface patches of reflection
\cite{moffat}:
   \begin{equation}
      h(t) = \sum_{i} h_i \delta(t-2r_i/c)\,,
   \end{equation}
where $h_i$ is the intensity of reflection from the bright points composing the radiolocation portrait,
$r_i$ are the locations of these points on the object.

The individual character of radiolocation portraits allows to use them for solving the pattern-recognition problem.

For the radial size $\sim 10$ m one has to ensure the distance resolution $\delta r \sim 0.5$ m (by comparison, the best resolution achieved at modern ground-based astronomical radars Goldstone and Aresibo is $\sim 4$ meters \cite{benner14}) which corresponds to duration of probing radar signal pulse $\sim 3.5$ ns.

Registration and processing of these signals are a matter of considerable difficulties due to the broad band of frequencies they occupy. Nevertheless, the periodic character of the signal permits the use of the stroboscopic effect in radio engineering, emerging upon strobing of the signals by a sequence of window functions with closely spaced repetition frequency. The procedure for the microwave signals can be realized in a balanced mixer when a pulse signal of a heterodyne repeating the probing signal is injecting to the reference channel.

   \begin{figure}
   \centering
      \includegraphics[width=1\textwidth]{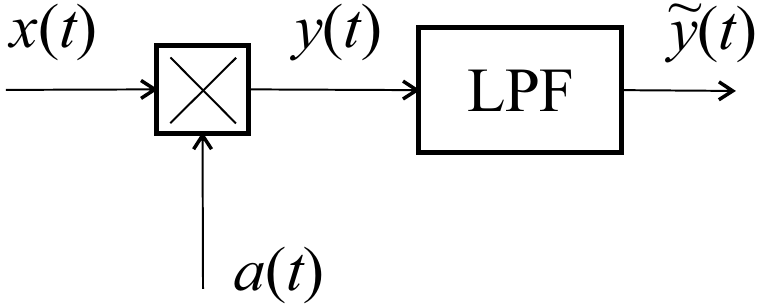} 
   \caption{The mathematical model of the stroboscopic transformer of the timescale.}
              \label{Fig2}
    \end{figure}

\section{Stroboscopic transformation of reflected signals}\label{transformation}

   \begin{figure}
   \centering
   \includegraphics[width=1\textwidth]{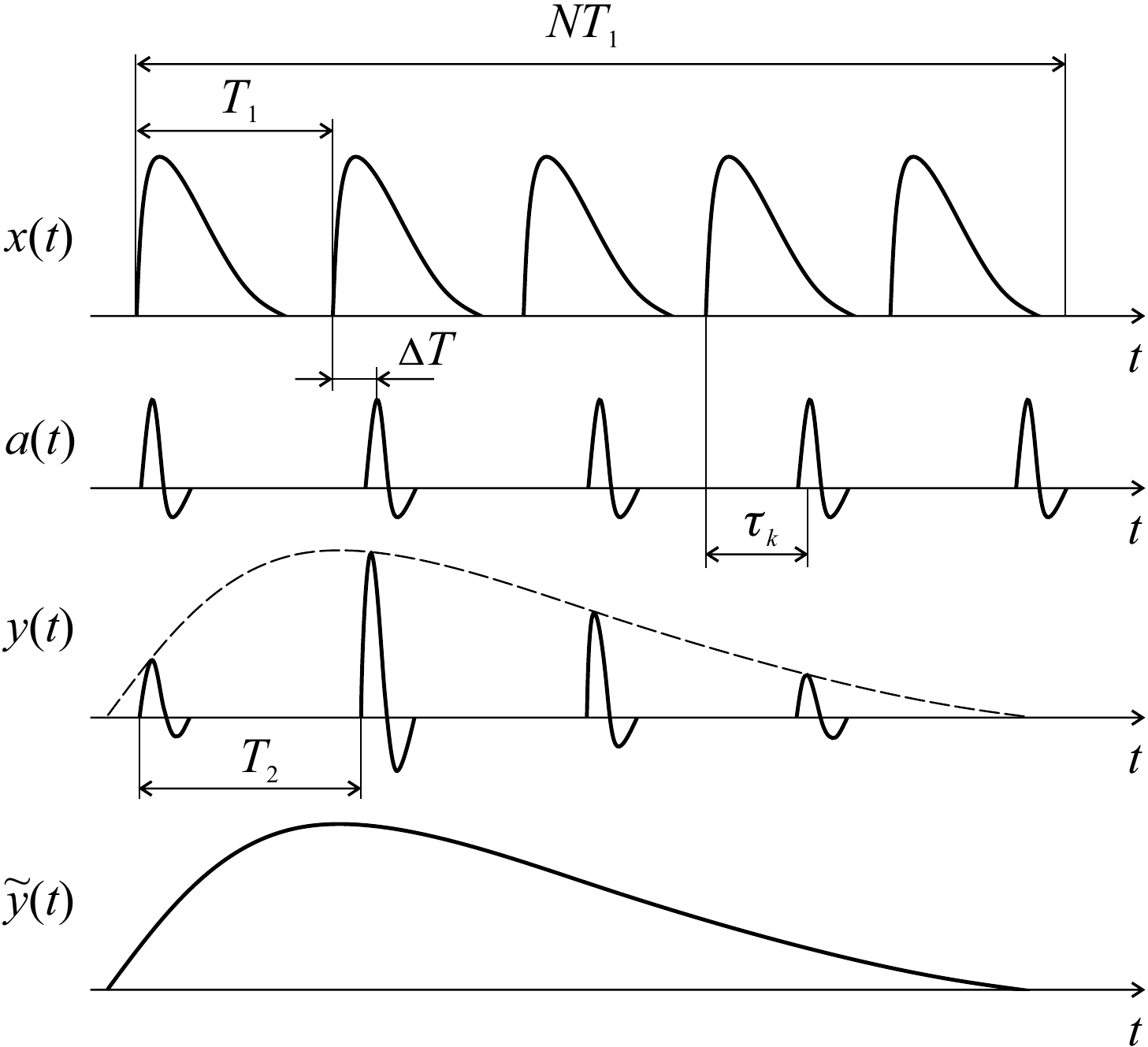}  
   \caption{The stroboscopic transformation of the timescale of periodic signals.}
              \label{Fig3}
    \end{figure}

Processing of reflected back ultrashort radio signals with wide frequency band can be achieved by  periodic repetition of the probing signal followed by stroboscopic transformation of the reflected signal’s time scale by $10^3 \div 10^5$ times.

Advantages of this method for our application are as follows:
- bulk of amplification comes in a narrow frequency band (few kHz), that  significantly simplifies the RF chain of receiver;
- the intermediate frequency (frequency shift between the carriers of probing and reflected signals) is formed naturally by the Doppler shift $\Omega$ when asteroid is moving towards the radar; this simplifies the receiver's hardware design (no need for additional high-frequency generator). Here $\Omega=2V_r/c \omega_0$, where $\omega_0$ is carrier frequency of probing signal, $V_r$ is radial velocity of the asteroid;
- no need to create auto-shift of RF pulse necessary for stroboscopic transformation because the required shift $\Delta T$ is formed due to relative motion of the asteroid and the radar. Here $\Delta T = 2V_r/c T$, where $T$ is the probing period;
- the transformed narrow-band signal can be transmitted to the base station for processing and analysis, if necessary.

The model of stroboscopic transformer which consists of a mixer multiplying the analyzed, $x(t)$, and the strobing $a(t)$ signals, and the low-pass filter (LPF) is presented in Fig.~\ref{Fig2}.

Fig.~\ref{Fig3} illustrates the principle of stroboscopic transformation where
   \begin{equation}\label{eq3}
      x(t) = \sum_{k=0}^{N} x_0(t-kT_1),  \qquad a(t) = \sum_{k=0}^{N} a_0(t-kT_2)
   \end{equation}
are the input and strobing signals of the stroboscopic transformer, $T_1$, $T_2$ are the periods of repetition of the signal and strobe, $\Delta T = T_2-T_1 =T_1/N$ is the reading period, $\tau_k=k\Delta T$ is the shift of the strobing pulse in the $k$-th period of the of the input signal, $N$ is the coefficient of the spectral transformation. Usually, $N\gg 1$ and ranges from $10^4$ to $10^6$.
The capabilities of the stroboscopic transformation are best implemented in the oscillographic equipment.

For radar probing the microwave mixer is used as a multiplier and the radio signals \eqref{eq3} have complex-valued representations
   \begin{equation}
      \dot{x}(t) = \sum_{k=0}^{N} \dot{A}_1(t-kT_1)e^{j\omega_1 t},  \qquad \dot{a}(t) = \sum_{k=0}^{N} \dot{A}(t-kT)e^{j\omega_0 t}\,,
   \end{equation}
where $\dot{A}_1(t)$,  $\dot{A}(t)$ are the complex modulation envelopes of the input and gating radio impulses, $\omega_0$ and $T$ are the carrier frequency and the period of repetition of the probing signal respectively.

The frequency shift $\Omega=\omega_1 - \omega_0 = 2\omega_0 V_r/c$ of the carrier frequency and the period decrease $T_1=T(1-2V_r/c)$ will take place in the reflected signal at the expense of the asteroid's high radial speed $V_r$.
This allows us to consider the circuit Fig.~\ref{Fig2} operating conditions as a radio impulse stroboscopic transformation \cite{zakharchenko11b} to low intermediate frequency $\Omega$, reading increment $\Delta T=2TV_r/c$ and the coefficient of the spectral transformation $N=T/\Delta T=c/2V_r$. The LPF in the circuit (Fig.~\ref{Fig2}) then should be replaced by the narrow-band tracking filter selecting the Doppler frequency $\Omega$.

The radial asteroid's velocity $V_r$ must be measured independently  by the narrow-band methods of finding the gravity centre of the reflected signal. One of the effective methods for the estimating the radial velocity of an asteroid in real time is the method of fractional differentiation of a Doppler signal \cite{zk14}.

The characteristic profile of the signal's spectrum on LPF entry is shown in Fig.~\ref{Fig4}.

The radiolocation system must possess the capability of reorganizing the repetition period $T$ for the probing signal in case of coincidence of the Doppler shift with the clock frequency of one of the harmonics and dependent on it spectral components.

In stroboscopic location it is convenient to choose of the same type the signal shapers of the probing, $\dot{s}(t)$, and strobing $\dot{a}(t)$ signals. It is assumed that the shapers operate both from the common collector of the carrier frequency $\omega_0$: $\dot{s}(t) \sim \dot{a}(t)$. In this case the radiolocation portrait of the object simulated by the aggregate of the bright points with consideration of the Doppler effect is described by the expression
   \begin{equation}
      \dot{x}(t) = \sum_{i}  h_i  e^{-j\omega \tau_i} \sum_{k=0}^{N} \dot{A}(t-2r_i/c-kT_1)e^{j\omega_1 t}\,.
   \end{equation}
The output signal of the radio impulse strobing circuit extracted by the filter in the vicinity of the frequency $\Omega$ in the case of the narrow probing strobing pulse is
asymptotically ($N\to\infty$) the narrow-band one \cite{zakharchenko09}
   \begin{equation}
      {y}(t) =\frac{1}{2T} {\rm Re}\left\{  e^{j\Omega t}\int_{-\infty}^{\infty}{\dot{A}_1(t^{\prime})A^*(t-\frac{t^{\prime}}{N})dt^{\prime}} \right\}
   \end{equation}
and with the use of the single-type signal conditioners $\dot{s}(t)$ and  $\dot{a}(t)$ is described by the relation
   \begin{equation}
      \dot{y}(t) \sim  \sum_{i} h_i \dot{B}\left(\frac{t}{N}-2r_i/c \right) e^{j(\Omega t+\varphi_i)}\,,
   \end{equation}
where $\dot{B}(t)=\frac{1}{2}\int_{-\infty}^{\infty}{\dot{A}(t^{\prime})A^*(t^{\prime}-t) dt^{\prime}}$ is the modulation envelope of the autocorrelation function of the probing signal which provides ultra-high resolution.

Thus, provided the resolution of individual bright points on the asteroid's surface, the modulation envelope of the output signal of the radio impulse strobing circuit describes the radiolocation portrait of the object in the transformed time scale.

\section{Estimate of asteroid's sizes}\label{Estimate}

   \begin{figure}
   \centering
   \includegraphics[width=1\textwidth]{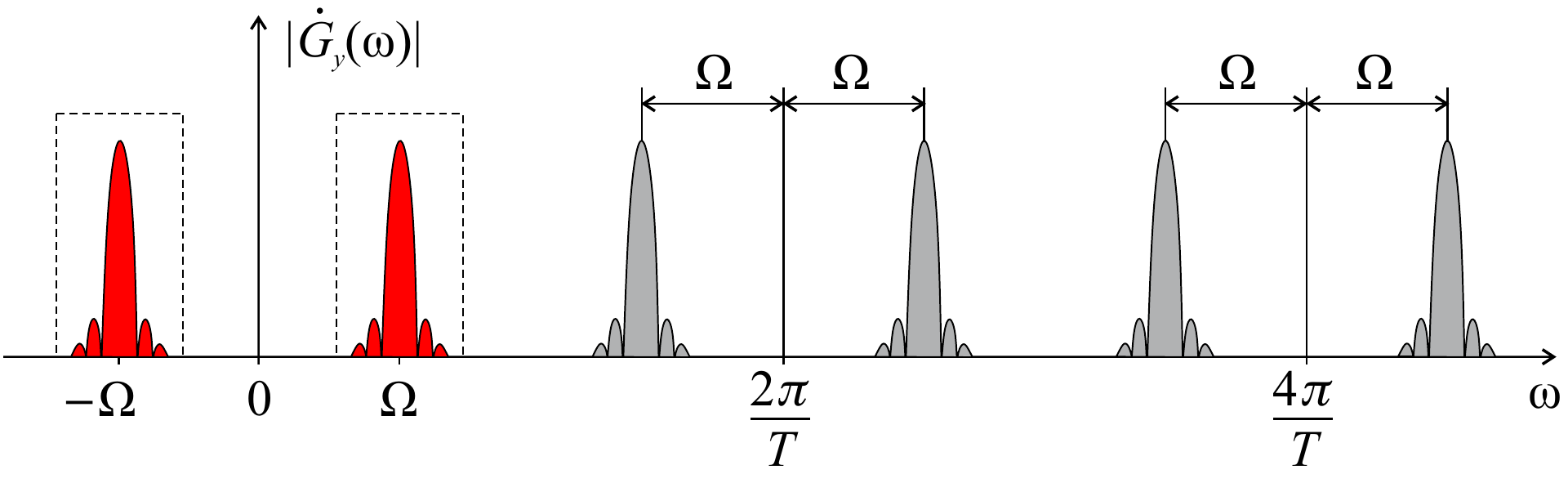} 
   \caption{The spectrum $|\dot{G}_y(\omega)|$ of the signal $y(t)$ on return of the stroboscopic mixer for $\Omega < \pi/T$. Red colour denotes the filtered spectral components.}
              \label{Fig4}
    \end{figure}

It is known that the characteristic feature of passive cosmic objects is their spin rotation due to drag-free flight \cite{Ostro97}. The typical periods of rotation range from hours to  minutes \cite{pravec02} whereas  the fastest spinning of all currently observed asteroids has the period of 24.5 seconds \cite{ryan14}.

The object surfaces reflecting the probing signal change their positions with asteroid's rotation. Measuring duration $\tau_m = 2\Delta R_m/c$ of the radiolocation portrait $x(t)$ at different aspect angles and averaging results of measurements one can get fairly accurate assessment of the mean radius of the cosmic object $\tilde{r}$
   \begin{equation}
      \tilde{r} \approx \frac{c}{2}\cdot \frac{1}{M} \sum_{m=1}^{M}\tau_m\,,
   \end{equation}
where $\tau_m$ is the radiolocation portrait duration at $m$-th measurement, $M$ is the number of measurements.
For periodic probing the value $M$ should be chosen from the condition $M=T_V /T$. Here $T_V$ is the asteroid's rotation period ($\sim 10\div 100$ min) that can be determined through repeatability of the radiolocation portrait, $F$ is the repetition rate of the probing signal that has to be chosen in such way as to fulfill the condition $M > 100\div 1000$. In presence of several rotational axes (tumbling) one should account for the largest period $T_V$.

Thus, to estimate the average size of the hazardous cosmic object one has to perform its probing by the periodic sequence of ultra-high-resolution signals of nanosecond duration and to get a radiolocation portrait by stroboscopic processing. Further one should chose the number $M$ determined by repeatability of the radiolocation portraits and corresponding to the number of aspect angles of the object within the rotational period $T_v$ or within the largest period of rotation in case of tumbling. With that, one performs multiple metering of radiolocation portraits duration $\tau_m$ $(m=1,2,..., M)$ of the illuminated side of the object. Then the measured values $\tau_m$ should be averaged over the number of measurements and then the mean object's radius is assessed by the half of average spatial length  of the signal of the radiolocation portrait $\tilde{r}\approx 0.5 c\langle \tau_m\rangle$ and the linear dimension $L\approx 2\tilde{r}$.

\section{Quantitative assessments}\label{Quantitative}

For the cosmic object measuring $50$ meters in diameter and the period of spin rotation of $\sim 30$ min the velocity spectrum width equals to $\sim 0.1$ m/s.
The range resolution $\delta r \sim 0.5$ m can be provided by the probing signal duration of $\sim 3$ ns in the X band ($f_0 \sim 10$ GHz) with the signal's frequency band $\Delta f_0\sim 300$ MHz. For the asteroid's radial velocity of $\sim 20$ km/s the Doppler frequency will then be $F_0 = 2 f_0 V_r/c \sim 1.2$ MHz.

The cutout of reflections on the line of sight, which is provided by the narrow antenna pattern, allows us to select the repetition rate of the signals $F\sim 100$ kHz (the signal's period  $\sim 10 \mu$s). The time of measurement of the radiolocation portrait by the stroboscopic method (storage time) will be  $T_{\Sigma}\sim 2$ ms,
the sampling increment is then $\Delta T \sim 1.3$ ns, the coefficient of the spectral transformation equals to $N=c/2V_r \sim 75000$ and the frequency band of the output signal is $\Delta F =\Delta f/N \sim 4$ kHz at the carrier frequency $\sim 1.2$ MHz.

Using ordinary technical means this signal can be discretized, digitally filtered out and transmitted through the narrow-band communication channels.

\section{Conclusions}\label{Conclusions}

   \begin{enumerate}
      \item The use of suggested method of stroboscopic location makes it possible to compose radiolocation portraits of small asteroids in the transformed timescale. This allows to estimate the dimensions of passive cosmic objects via duration of their radiolocation portraits with distance-independent accuracy. With that, the realization of the method does not require the special autoshift circuit which is used in the stroboscopic oscillographs because the strobe shift within the period is performed at the expense of the Doppler effect for the repetition frequency, which is usually neglected in radiolocation.
      \item The method of stroboscopic location which makes it possible to estimate the longitudinal asteroid’s dimension serves as the complementary to the traditional methods used in the modern radar astronomy which aimed at determination of the cross sectional dimensions.
      \item We plan to back the framework concept expressed in the present paper with proof by developing a 3D computer model of a the reflecting surface of a rotating asteroid.
   \end{enumerate}

\bigskip\medskip\noindent {\bf Acknowledgements}
\medskip

We thank Vitaly Korolev for assistance with vectorizing the figures and Vladimir Levi for revision of the language. This work was supported by the grants 13-01-97041r-povolzhie-a, 14-02-97001r-povolzhie-a from Russian Foundation for Basic Research.

This research was supported in part by the Russian Foundation for Basic Research under grants 13-01-97041r-povolzhie-a and 14-02-97001r-povolzhie-a.

\end{document}